# Electronic phonon induced magnetism in moiré Mott-Wigner crystals

Weide Liang[1†], Xiaoying Du[1†], Na Zhang[1], Hongyi Yu[1,2*]

[1] Guangdong Provincial Key Laboratory of Quantum Metrology and Sensing & School of Physics and Astronomy, Sun Yat-Sen University (Zhuhai Campus), Zhuhai 519082, China

[2] State Key Laboratory of Optoelectronic Materials and Technologies, Sun Yat-Sen University (Guangzhou Campus), Guangzhou 510275, China

† These authors contributed equally to this work.

* E-mail: yuhy33@mail.sysu.edu.cn

**Abstract:** We show that magnetism in moiré Mott-Wigner crystals can be induced by the collective vibration of electrons around their equilibrium positions (namely electronic phonons), even without the electron-electron spin interaction. Due to a geometric valley-orbit coupling from the Berry phase effect, the zero-point energy of electronic phonons reaches minimum when electrons are fully valley polarized. This leads to a spontaneous magnetization when below a critical temperature. We further propose engineering magnetism in Mott-Wigner crystals through the photoexcitation of chiral electronic phonons.

The formation of periodic moiré patterns in van der Waals stacking of layered two-dimensional (2D) materials has introduced a new platform for studying exotic quantum states [1,2]. A variety of correlated electron crystals, including Mott insulators under integer fillings and generalized Wigner crystals under fractional fillings in moiré systems, as well as monolayer/bilayer Wigner crystals without the moiré pattern, have been detected in layered transition metal dichalcogenides (TMDs) [3-16]. Generally, it is expected that magnetism induced by the short-ranged spin interaction can emerge in these Mott-Wigner crystals (MWC), with the corresponding magnetic phase diagrams investigated in a series of theoretical works [17-22]. However, experimental signatures for the spontaneous magnetism in MWC have remained elusive, which could be due to the weak spin interaction strength from the large inter-site separation combined with the moiré-trapping induced strong localization. Alternate mechanisms for engineering magnetism thus become important. For instance, a super-exchange interaction between moiré-trapped electrons can be mediated by photoexcited excitons [23-25] or doped electrons/holes [26-29]. Meanwhile, these quasiparticles introduce additional carriers to the system thus somehow change its insulating nature.

The emergence of moiré MWC under fractional fillings signifies the important role of inter-site Coulomb interactions. This gives rise to the collective vibration of electrons near their equilibrium positions, namely electronic phonons (EP), which has been studied in a series of theoretical works [30-34]. In GaAs-based quantum wells under high magnetic fields, transport signals indicating the existence of EP in Wigner crystals have been detected [35,36]. Very recently, several experiments reported the optical signatures of EP in TMDs systems under zero magnetic field and the EP-exciton interaction [37-39]. These findings suggest EP as a commonly existing intrinsic charge excitation in moiré MWC. As a nonlocal quasiparticle, the synchronized collective vibration in the EP mode can potentially mediate the spin correlation between remote electrons, thus affects the magnetic properties of MWC.

In this work, we show that EP can introduce magnetism to moiré MWC, through a geometric valley-orbit coupling effect proportional to the electron Berry curvature. We show that EP energies vary with the valley configuration of MWC. The corresponding zero-point energy reaches its minimum when electrons are fully valley polarized, implying the emergence of ferromagnetism under a low temperature even when the electron spin interaction completely vanishes. We further propose engineering magnetism in MWC through a photoexcited nonequilibrium distribution of chiral EP.

We consider a system of 2D massive Dirac electrons subjected to a moiré potential and strong many-body Coulomb interactions. The Hamiltonian is $\hat{H} = \sum_n \big(\hat{H}_n + V_{\text{moiré}}(\mathbf{r}_n) + \frac{1}{2}\sum_{n' \neq n} V(\mathbf{r}_n - \mathbf{r}_{n'})\big)$. Here $\hat{H}_n = \sum_{\tau_n} \left(v_F \tau_n \hat{p}_{n,x}\hat{\sigma}_{n,x} + v_F \hat{p}_{n,y}\hat{\sigma}_{n,y} + \frac{\Delta}{2}\hat{\sigma}_{n,z}\right) \otimes |\tau_n\rangle\langle\tau_n|$ is the massive Dirac Hamiltonian of $n$-th electron with $\tau_n = \pm 1$ the valley pseudospin, $v_F$ the Fermi velocity, $\hat{p}_{n,x/y}$ the in-plane momentum operators and $\hat{\sigma}_{n,x/y/z}$ the 2x2 Pauli matrices [40]. $V_{\text{moiré}}(\mathbf{r}_n)$ is the moiré potential experienced by $n$-th electron with spatial coordinate $\mathbf{r}_n$, and $V(\mathbf{r}_n - \mathbf{r}_{n'})$ is the Coulomb interaction. By projecting the Hamiltonian to the conduction band subspace [41], it can be rewritten as $\hat{H} \approx \hat{H}_{\text{EP}} + \hat{H}_{\text{BG}}$, with

$$\begin{aligned} \hat{H}_{\text{EP}} &= \sum_n \left( -\frac{\hbar^2}{2m}\frac{\partial^2}{\partial \mathbf{r}_n^2} + V_{\text{moiré}}(\mathbf{r}_n) + \frac{1}{2}\sum_{n' \neq n} V(\mathbf{r}_n - \mathbf{r}_{n'}) \right), \\ \hat{H}_{\text{BG}} &= \sum_n \left( \frac{\Omega}{4}\frac{\partial^2 U_n}{\partial \mathbf{r}_n^2} - \sum_{\tau_n} \frac{\tau_n \Omega}{2}\left(\frac{\partial U_n}{\partial \mathbf{r}_n} \times \hat{\mathbf{p}}_n\right)_z \otimes |\tau_n\rangle\langle\tau_n| \right). \end{aligned} \tag{1}$$

Here $\hat{H}_{\text{EP}}$ is the standard many-body Hamiltonian of electrons with an effective mass $m \equiv \frac{\hbar^2 \Delta}{2v_F^2}$ under the moiré potential. $\hat{H}_{\text{BG}}$ is the lowest-order correction from the band geometry or Berry phase effect, which consists of a Darwin term $\frac{\Omega}{4}\partial^2 U_n/\partial \mathbf{r}_n^2$ and an Ising-type valley-orbit coupling (VOC) term $\frac{\tau_n \Omega}{2}(\partial U_n/\partial \mathbf{r}_n \times \hat{\mathbf{p}}_n)_z$. Here $\tau_n \Omega \equiv \frac{2v_F^2}{\Delta^2}\tau_n$ is the Berry curvature in $\tau_n \mathbf{K}$ valley and $U_n \equiv V_{\text{moiré}}(\mathbf{r}_n) + \sum_{n' \neq n} V(\mathbf{r}_n - \mathbf{r}_{n'})$ is the total potential experienced by $n$-th electron. Note that the strength $\frac{\Omega}{2}$ of this VOC is six orders larger than $\frac{\hbar^2}{4m_0^2 c^2}$ of the relativistic spin-orbit coupling (SOC) with $m_0$ the free electron mass, thus can be significant even when electrons are separated by several nm. The VOC term doesn't flip the valley pseudospin, but introduces a valley-dependent energy correction. It can be intuitively viewed as the effect of the Berry curvature as an out-of-plane momentum-space magnetic field, which has been found to result in a splitting between $2p_\pm$ Rydberg excitons [42-45]. In monolayer TMDs, the valley pseudospin of band edge carriers is locked to the spin due to a large SOC splitting [46]. Meanwhile, the Berry phase effect gives rise to an anomalous contribution to the valley magnetic moment which is $\tau_n \frac{e}{\hbar}\frac{v_F^2}{\Delta}$ under the massive Dirac model [47]. This anomalous magnetic moment has a magnitude reaching $\sim 2\mu_{\mathrm{B}}$ in monolayer TMDs and can be even larger in gapped graphene with smaller band gaps. In systems with negligible SOC splitting so the spin expectation value is zero, the valley magnetism can originate from this anomalous magnetic moment. Below we shall view the valley pseudospin of electrons simply as the spin.

In the strong correlation limit and when the filling factor $v$ (defined as the average electron number per moiré unit cell) corresponds to an integer or certain fractional values, the system described by $\widehat{H}_{\mathrm{EP}}$ forms an MWC. Fig. 1(a) schematically shows a triangular moiré pattern, where a triangular (honeycomb) MWC can form under the filling $v = 1$ ($v = 2/3$). In this work, we assume that each electron is strongly localized near a moiré site $\mathbf{R}_n$ with negligible inter-site hopping. $\widehat{H}_{\mathrm{EP}}$ then describes the collective vibration of electrons near moiré sites, giving rise to a series of EP modes labeled by the branch index $l$ and wave vector $\mathbf{k}$. To solve these modes, we follow the standard treatment for the phonon problem [32]. After expanding $V_{\mathrm{moiré}}$ and $V$ up to the second order of $\mathbf{r}_n - \mathbf{R}_n$ and introducing the EP annihilation (creation) operator $\hat{a}_{l,\mathbf{k}}$ ($\hat{a}_{l,\mathbf{k}}^\dagger$), we can write $\widehat{H}_{\mathrm{EP}} = \sum_{l,\mathbf{k}} \omega_{l,\mathbf{k}} \left(\hat{a}_{l,\mathbf{k}}^\dagger \hat{a}_{l,\mathbf{k}} + \frac{1}{2}\right)$ [41]. Here $\omega_{l,\mathbf{k}}$ is the EP energy, and the corresponding in-plane polarization vector is denoted as $\boldsymbol{\epsilon}_{l,\mathbf{k}}$. We show in Fig. 1(b) and 1(c) the calculated EP dispersions for the triangular and honeycomb MWC under $v = 1$ and 2/3, respectively. Here a Rytova-Keldysh Coulomb potential form $V(\mathbf{r}) \equiv \frac{\pi}{2r_0}\left[H_0\left(\frac{r}{r_0}\right) - Y_0\left(\frac{r}{r_0}\right)\right]$ has been used with $H_0$ ($Y_0$) the Struve (2nd-kind Bessel) function and $r_0 \approx 4.5$ nm the 2D screening length of the TMDs monolayer. The other parameters are given in the figure caption. The EP energies are generally in the order of several tens meV, which comes from the small effective mass of the electron and the weakly screened Coulomb interaction in atomically-thin 2D materials. Note that the inversion ($\mathcal{P}$) and time-reversal ($\mathcal{T}$) symmetries are conserved in both the triangular and honeycomb MWC described by $\widehat{H}_{\mathrm{EP}}$. As a result, electron motions of EP modes in triangular MWC are linearly polarized. On the other hand, EP modes in the honeycomb crystal exhibit circularly polarized motions at the moiré Brillouin zone corners $\pm\mathbf{K}_{\mathrm{m}}$, with opposite polarizations for different sublattices, see Fig. 1(d). Meanwhile, the EP chirality defined as the expectation value of the total angular momentum $\hat{L}_z \equiv \sum_n (\mathbf{r}_n \times \hat{\mathbf{p}}_n)_z$ is zero in both triangular and honeycomb MWC.

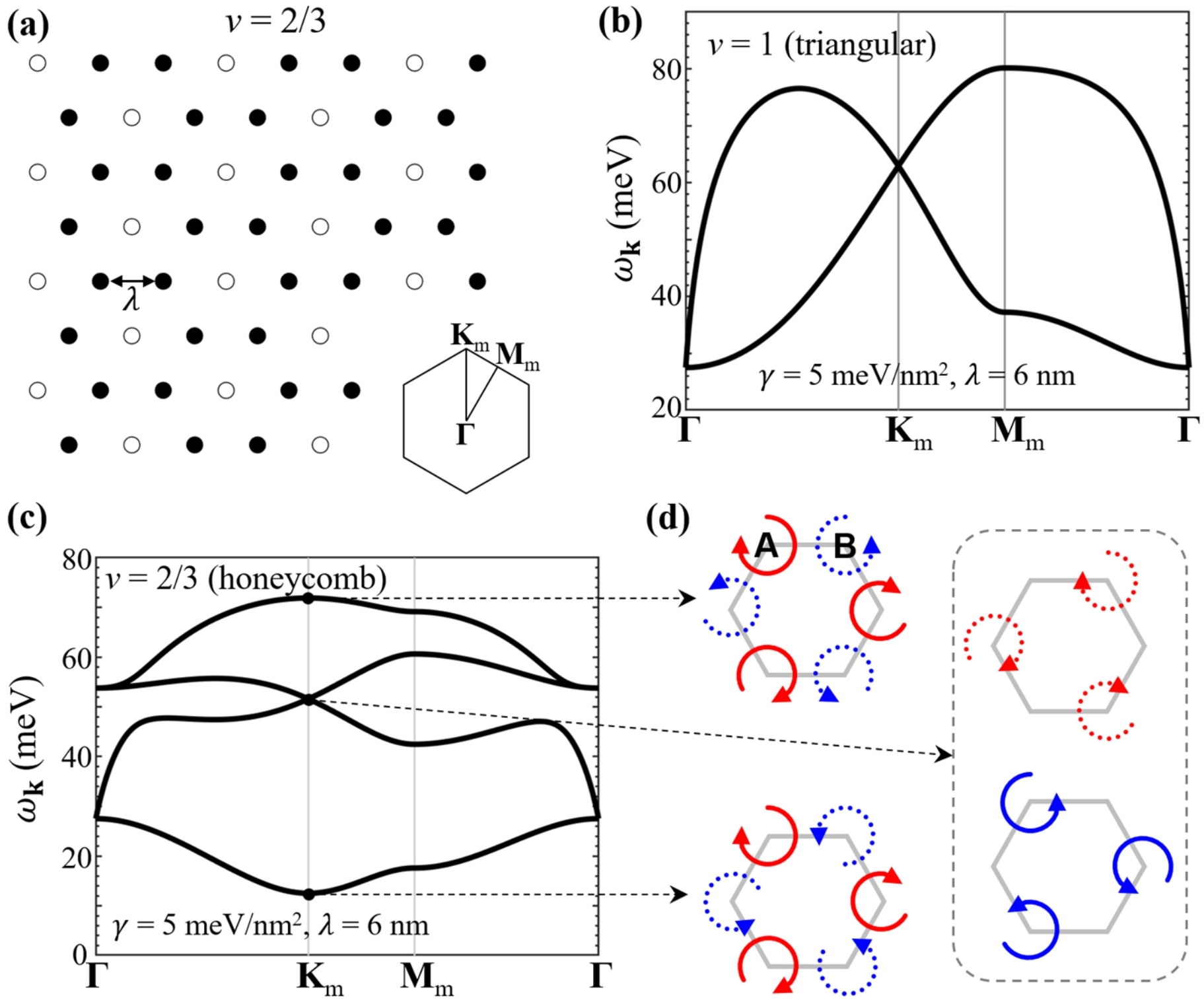

**Figure 1.** (a) A schematic illustration of the honeycomb MWC under a filling factor $\nu = 2/3$, formed in a triangular moiré pattern with a wavelength $\lambda$. Solid (empty) dots denote filled (unfilled) moiré potential minima. The inset shows the moiré Brillouin zone. (b) The calculated EP dispersion of the triangular MWC under $\nu = 1$, without the band geometry correction. The system parameters are set as: effective mass $m = 0.5m_0$, moiré wavelength $\lambda = 6$ nm and moiré confinement strength $\gamma \equiv \nabla^2 V_{\text{moiré}}(\mathbf{r})|_{\mathbf{r}=\mathbf{R}_n} = 5$ meV/nm$^2$. (c) The EP dispersion of the honeycomb MWC under $\nu = 2/3$. The other parameters are the same as in (b). (d) The circularly polarized electron motions of the four EP modes at $\mathbf{K}_{\mathrm{m}}$ of the honeycomb MWC. A and B are the two sublattices. Red (blue) color denotes the left-handed (right-handed) circular polarization.

In the band geometry part $\hat{H}_{\mathrm{BG}}$, the Darwin term only gives a constant energy shift thus can be dropped. The VOC term, describing the coupling of electron's circular motion to a momentum-space magnetic field from the Berry curvature (see the illustration in Fig. 2(a)), can be rewritten as the sum of $t_{\mathbf{q}}\hat{a}^{\dagger}_{l,\mathbf{k}}\hat{a}_{l',\mathbf{k}-\mathbf{q}}$, $t_{\mathbf{q}}\hat{a}^{\dagger}_{l,\mathbf{k}}\hat{a}^{\dagger}_{l',\mathbf{q}-\mathbf{k}}$ and $t_{\mathbf{q}}\hat{a}_{l,-\mathbf{k}}\hat{a}_{l',\mathbf{k}-\mathbf{q}}$ with $t_{\mathbf{q}} \equiv \frac{1}{\sqrt{N}}\sum_n \tau_n e^{-i\mathbf{q}\cdot\mathbf{R}_n}$. Here we assume the complex coefficients of these terms with $\mathbf{q} \neq 0$ average out, since the phase of $t_{\mathbf{q}}$ varies with the spin configuration of all electrons (the effect of $\mathbf{q} \neq 0$ terms has been evaluated in systems with small site numbers [41], and it is found that the qualitative results remain the same). We then get

$$\hat{H}_{\mathrm{BG}} \approx m\tau\Omega \sum_{ll'\mathbf{k}} P_{ll',\mathbf{k}} \sqrt{\omega_{l,\mathbf{k}}\omega_{l',\mathbf{k}}} \left( \frac{\omega_{l,\mathbf{k}}+\omega_{l',\mathbf{k}}}{4} \hat{a}^{\dagger}_{l,\mathbf{k}}\hat{a}_{l',\mathbf{k}} - \frac{\omega_{l,\mathbf{k}}-\omega_{l',\mathbf{k}}}{8} \left( \hat{a}^{\dagger}_{l,\mathbf{k}}\hat{a}^{\dagger}_{l',-\mathbf{k}} - \hat{a}_{l,-\mathbf{k}}\hat{a}_{l',\mathbf{k}} \right) \right). \quad (2)$$

Here $P_{ll',\mathbf{k}} \equiv i\left(\boldsymbol{\epsilon}_{l,\mathbf{k}} \times \boldsymbol{\epsilon}^{*}_{l',\mathbf{k}}\right)_z$ and $\tau \equiv \frac{1}{N}\sum_n \tau_n$ is the average magnetic moment of MWC. The total Hamiltonian of MWC can be diagonalized by a Bogoliubov transformation, resulting in

$$\hat{H} \approx \sum_{l,\mathbf{k}} \tilde{\omega}_{l,\mathbf{k}}(\tau)\left(\hat{b}_{l,\mathbf{k}}^{\dagger}\hat{b}_{l,\mathbf{k}} + \frac{1}{2}\right). \tag{3}$$

The energy $\tilde{\omega}_{l,\mathbf{k}}(\tau)$ and polarization of the valley-orbit coupled EP are generally different from those of the original EP. Note that in honeycomb MWC with two sublattices, $\tau P_{ll',\mathbf{k}}$ should be replaced by $\tau_{\mathrm{A}} P_{ll',\mathbf{k}}^{(\mathrm{A})} + \tau_{\mathrm{B}} P_{ll',\mathbf{k}}^{(\mathrm{B})}$, where $\tau_{\mu}$ is the average magnetic moment and $P_{ll',\mathbf{k}}^{(\mu)} \equiv i\left(\boldsymbol{\epsilon}_{l,\mathbf{k}}^{(\mu)} \times \boldsymbol{\epsilon}_{l',\mathbf{k}}^{(\mu)*}\right)_z$ with $\boldsymbol{\epsilon}_{l,\mathbf{k}}^{(\mu)}$ the polarization vector of sublattice $\mu$ = A, B. In this case, the EP energy $\tilde{\omega}_{l,\mathbf{k}}(\tau_{\mathrm{A}}, \tau_{\mathrm{B}})$ varies with both $\tau_{\mathrm{A}}$ and $\tau_{\mathrm{B}}$.

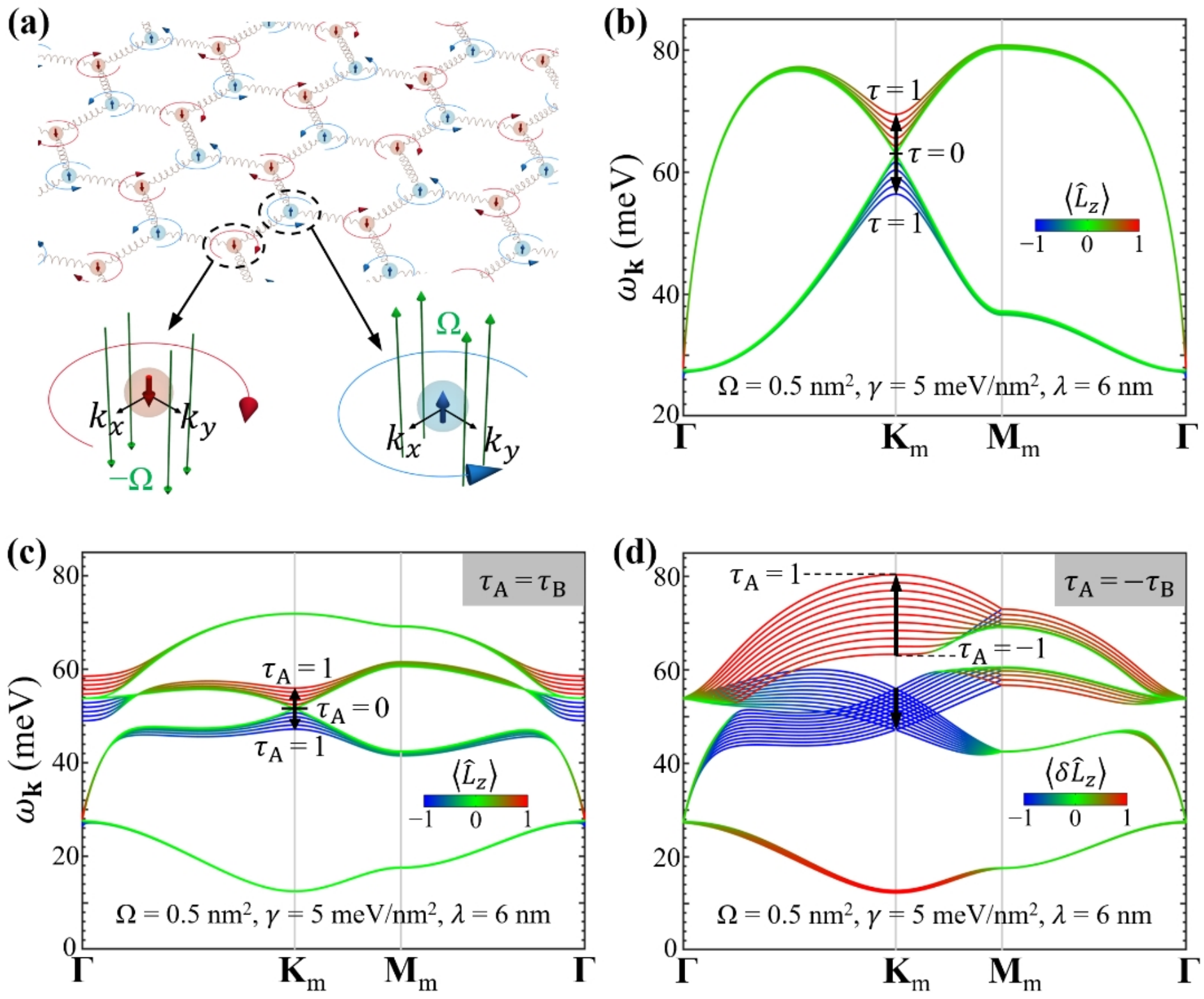

**Figure 2.** (a) A schematic illustration of the coupling between Berry curvatures and circular motions of electrons in a honeycomb MWC. The Berry curvature $\pm\Omega$ (thin green arrows), whose sign is locked to the electron spin/valley (thick red and blue arrows), acts as a momentum-space magnetic field. The electron's circular motion thus shifts the energy of EP. (b) The EP dispersion of a triangular MWC under $\Omega = 0.5$ nm$^2$ and other parameters the same as in Fig. 1(b). Different lines correspond to EP dispersions under different average magnetic moment $\tau$, with black arrows showing the evolution when $\tau$ is increased from 0 to 1. The line color corresponds to the EP chirality $\langle \hat{L}_z \rangle$. (c) The EP dispersion of a ferromagnetic honeycomb MWC with $\tau_{\mathrm{A}} = \tau_{\mathrm{B}}$ under $\Omega$ = 0.5 nm$^2$. (d) The EP dispersion of an antiferromagnetic honeycomb MWC with $\tau_{\mathrm{A}} = -\tau_{\mathrm{B}}$. The line color corresponds to the angular momentum difference $\langle \delta\hat{L}_z \rangle$ between A and B sublattices. The black arrows show the evolution of EP dispersion when $\tau_{\mathrm{A}}$ is increased from −1 to 1. Other system parameters in (c) and (d) are the same as those in Fig. 1(c).

Now we investigate how EP responds to the magnetism in MWC. Fig. 2(b) shows the EP dispersion of a triangular MWC under various average magnetic moments. Without magnetism ($\tau$ = 0), $\mathcal{P}$ and $\mathcal{T}$ symmetries restrict the electron motions to be linearly polarized with zero chirality for all EP modes, and result in the doubly degeneracies at $\mathbf{\Gamma}$ and $\mathbf{K}_{\mathrm{m}}$. The ferromagnetism breaks $\mathcal{T}$ symmetry and introduces finite energy splittings at $\mathbf{\Gamma}$ and $\mathbf{K}_{\mathrm{m}}$, where EP modes exhibit ±1 chirality values. Fig. 2(c) shows the EP dispersions of a ferromagnetic honeycomb MWC ($\tau_{\mathrm{A}} = \tau_{\mathrm{B}}$), with the line color indicating the chirality. It also shows degeneracy liftings and ±1 chirality values at $\mathbf{\Gamma}$ and $\mathbf{K}_{\mathrm{m}}$. Note that the ferromagnetism in Fig. 2(b,c) conserves $\mathcal{P}$ symmetry, thus EP modes $\hat{b}_{l,\mathbf{k}}^{\dagger}$ and $\hat{b}_{l,-\mathbf{k}}^{\dagger}$ exhibit the same energy and chirality. On the other hand, the antiferromagnetic honeycomb MWC with $\tau_{\mathrm{A}} = -\tau_{\mathrm{B}}$ breaks both $\mathcal{P}$ and $\mathcal{T}$ but converses their combination, which leads to doubly degeneracies and zero chirality at $\mathbf{\Gamma}$ and $\mathbf{K}_{\mathrm{m}}$, see Fig. 2(d). We introduce $\delta\hat{L}_z \equiv \sum_{n\in\mathrm{A}}(\mathbf{r}_n \times \hat{\mathbf{p}}_n)_z - \sum_{n\in\mathrm{B}}(\mathbf{r}_n \times \hat{\mathbf{p}}_n)_z$ as the angular momentum difference between A and B sublattices, whose expectation values are shown as colors of the lines in Fig. 2(d). The expectation value of $\delta\hat{L}_z$ is close to ±1 in a large area of moiré Brillouin zone due the opposite circular motions of A and B sublattices, giving rise to significant EP energy shifts with the magnetic moment (see the illustration in Fig. 2(a)).

Below we show how magnetism emerges spontaneously under a low-enough temperature, even when the electron spin interaction completely vanishes. Since the EP energies are in the order of ~ 10 meV, we consider a much lower temperature which results in negligible EP occupations. In the thermal equilibrium of a triangular MWC with $N$ electrons, the probability to obtain an average magnetic moment $\tau$ is $P(\tau) \propto C_{N(\tau+1)/2}^{N} \exp\left(\frac{N\Delta E_0(\tau)}{k_B T}\right)$. Here $C_{N(\tau+1)/2}^{N} \equiv \frac{N!}{\left(\frac{N}{2}+\frac{N}{2}\tau\right)!\left(\frac{N}{2}-\frac{N}{2}\tau\right)!}$, and $\Delta E_0(\tau) \equiv \frac{1}{2N}\sum_{l,\mathbf{k}} \omega_{l,\mathbf{k}} - \frac{1}{2N}\sum_{l,\mathbf{k}} \widetilde{\omega}_{l,\mathbf{k}}(\tau)$ is the band geometry induced correction to the zero-point energy. The function form of $\Delta E_0(\tau)$ determines when the spin distribution becomes peaked at a finite magnetic moment, which signals a spontaneous magnetization. The calculated zero-point energy correction as a function of $\Omega\tau$ is shown in Fig. 3(a) inset, which is found to be positive and generally much smaller than the EP energy. In fact, due to the symmetries of MWC, the zero-point energy is an even function of $\tau$ which can then be expanded as $\Delta E_0(\tau) \approx \alpha\Omega^2\tau^2$. The fitting in Fig. 3(a) inset gives $\alpha$ = 0.10 meV/nm$^4$ under the adopted system parameters. Fig. 3(a) shows the spin distribution of a triangular MWC. When above a critical temperature $T_c^{(t)} \equiv 2\alpha\Omega^2 k_B^{-1}$, the distribution exhibits a single-peaked shape centered at $\tau = 0$, whose width broadens with the lowering of temperature. When below this critical value, the spin distribution becomes double-peaked, and the peak positions quickly shift to $\tau \approx \pm 1$ if further lowering the temperature. This corresponds to a phase transition from paramagnetic to ferromagnetic state. The ferromagnetism quantified by the thermally-averaged spin structure factor $S_{\mathbf{k}=0} \equiv N\langle\tau^2\rangle_T$ is shown in Fig. 3(b), indicating a sharp transition from ~1 when above the critical temperature to ~$N$ when below it.

For the honeycomb MWC with both A and B sublattices containing $N$ electrons, the spin distribution under a low temperature is $P(\tau_{\mathrm{A}},\tau_{\mathrm{B}}) \propto C^{N}_{\frac{N}{2}(\tau_A+1)} C^{N}_{\frac{N}{2}(\tau_{\mathrm{B}}+1)} e^{\frac{N\Delta E_0(\tau_{\mathrm{A}},\tau_{\mathrm{B}})}{k_B T}}$. The symmetry constraint requires $\Delta E_0(\tau_{\mathrm{A}},\tau_{\mathrm{B}}) = \Delta E_0(\tau_{\mathrm{B}},\tau_{\mathrm{A}}) = \Delta E_0(-\tau_{\mathrm{A}},-\tau_{\mathrm{B}})$, therefore it can be expanded as $\Delta E_0(\tau_{\mathrm{A}},\tau_{\mathrm{B}}) \approx \alpha_+\Omega^2\left(\frac{\tau_{\mathrm{A}}+\tau_{\mathrm{B}}}{2}\right)^2 + \alpha_-\Omega^2\left(\frac{\tau_{\mathrm{A}}-\tau_{\mathrm{B}}}{2}\right)^2$ which agrees excellently with the calculated value (see Fig. 3(c)). As shown in Fig. 3(d-f), the spin distribution peak position is shifted from zero when the temperature is above the critical value $T_{\mathrm{c}}^{(\mathrm{h})} \equiv k_B^{-1}\max(\alpha_+\Omega^2, \alpha_-\Omega^2)$ to two finite and opposite magnetic moments when below it, again showing a magnetic phase transition. Below this critical temperature, the honeycomb MWC exhibits $\tau_{\mathrm{A}} = \tau_{\mathrm{B}} \approx \pm 1$ (ferromagnetism) when $\alpha_+ > \alpha_-$, or $\tau_{\mathrm{A}} = -\tau_{\mathrm{B}} \approx \pm 1$ (antiferromagnetism) when $\alpha_+ < \alpha_-$. The fitting in Fig. 3(c) gives $\alpha_+ = 0.10$ meV/nm$^4$ $> \alpha_- = 0.02$ meV/nm$^4$, implying that the system tends to form an out-of-plane ferromagnetic phase when below the critical temperature. Fig. 3(g) shows the thermally-averaged spin structure factors $S^{\mathrm{AA}}_{\mathbf{k}=0} \equiv N\langle\tau_{\mathrm{A}}^2\rangle_T$ and $S^{\mathrm{AB}}_{\mathbf{k}=0} \equiv N\langle\tau_{\mathrm{A}}\tau_{\mathrm{B}}\rangle_T$ of the honeycomb MWC. $S^{\mathrm{AA}}_{\mathbf{k}=0}$ is rather similar to $S_{\mathbf{k}=0}$ of the triangular MWC, whereas $S^{\mathrm{AB}}_{\mathbf{k}=0}$ is close to zero when the temperature is much higher than the critical value but increases to $S^{\mathrm{AB}}_{\mathbf{k}=0} \approx S^{\mathrm{AA}}_{\mathbf{k}=0}$ when near it.

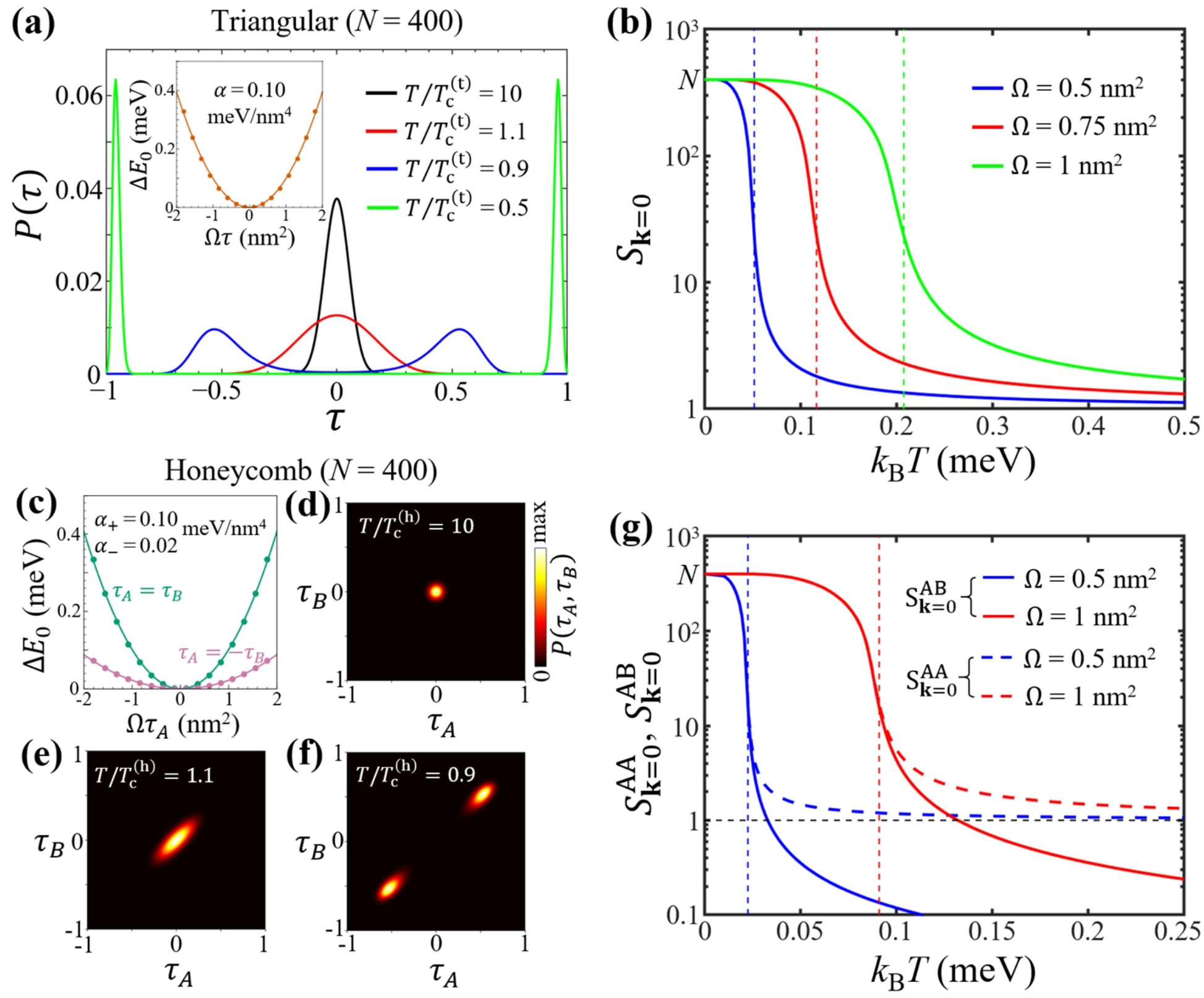

**Figure 3.** (a) The spin distribution $P(\tau)$ in a triangular MWC with $N = 400$ sites under several different temperatures. Other parameters are the same as in Fig. 2(b). $P(\tau)$ is peaked at an average magnetic moment $\tau = 0$ ($\tau \neq 0$) when $T > T_{\mathrm{c}}^{(\mathrm{t})} \equiv 2\alpha\Omega^2/k_B$ ($T < T_{\mathrm{c}}^{(\mathrm{t})}$). The inset is the calculated zero-point energy correction $\Delta E_0(\tau)$ as a function of $\Omega\tau$, showing a parabolic form $\Delta E_0(\tau) = \alpha\Omega^2\tau^2$ with $\alpha = 0.10$ meV/nm$^4$. (b) The thermally-averaged spin structure factor $S_{\mathbf{k}=0}$ as a function of temperature. Vertical dashed lines mark the critical temperatures $T_{\mathrm{c}}^{(\mathrm{t})}$. Different colors correspond to different Berry curvature

values. (c) $\Delta E_0(\tau_A,\tau_B)$ as a function of $\Omega(\tau_A \pm \tau_B)$ for the honeycomb MWC in Fig. 2(c,d). (d-f) The spin distribution $P(\tau_A,\tau_B)$ under three different temperatures. (g) The thermally-averaged spin structure factors $S_{\mathbf{k}=0}^{\mathrm{AA}}$ and $S_{\mathbf{k}=0}^{\mathrm{AB}}$ of the honeycomb MWC. The vertical dashed lines mark the critical temperatures $T_{\mathrm{c}}^{(\mathrm{h})}$.

Under a temperature much higher than $T_{\mathrm{c}}^{(\mathrm{t/h})}$ but still below the MWC's melting temperature, no magnetism is expected to occur under thermal equilibrium. Meanwhile, a nonequilibrium distribution of EP can be dynamically established through photoexcitation processes like the infrared absorption and Raman scattering. Similar to recent experiments and proposals for manipulating ferromagnetic Chern insulators through optically injected spin-polarized holes [48-51], these photoexcited EP can also give rise to magnetism. Considering that the monolayer exciton has a large optical absorption and interacts strongly with EP [37-39], a stimulated Raman process can generate an EP pair of opposite wave vectors through the scattering with excitons. We set the laser polarization to be $\sigma_-$ ($\sigma_+$) for the pump (Stokes) laser, and write the laser-exciton and exciton-EP interaction Hamiltonian as [41]

$$\begin{aligned}\hat{H}_{\mathrm{int}} = &\sum_{l,\mathbf{k}}\left(\hat{b}_{l,\mathbf{k}} + \hat{b}_{l,-\mathbf{k}}^{\dagger}\right)\sum_{\mathbf{Q},\mathbf{G}}\sum_{\tau=\pm}\alpha_{l,\mathbf{k}+\mathbf{G}}\left|X_{\tau,\mathbf{Q}+\mathbf{k}+\mathbf{G}}\right\rangle\left\langle X_{\tau,\mathbf{Q}}\right| \\ &+\left(\gamma_{\mathrm{p}}e^{-i\omega_{\mathrm{p}}t}\left|X_{-,0}\right\rangle\langle\mathbf{0}| + \gamma_{\mathrm{s}}e^{-i\omega_{\mathrm{s}}t}\left|X_{+,0}\right\rangle\langle\mathbf{0}| + \mathrm{h.c.}\right).\end{aligned} \tag{4}$$

Here $|X_{\tau,\mathbf{Q}}\rangle$ is a $\tau\mathbf{K}$ valley exciton with a center-of-mass momentum $\mathbf{Q}$, $\alpha_{l,\mathbf{k}+\mathbf{G}}$ is the exciton-EP interaction strength with $\mathbf{G}$ the moiré reciprocal lattice vector, $\omega_{\mathrm{p}}$ ($\omega_{\mathrm{s}}$) is the pump (Stokes) laser frequency, $\gamma_{\mathrm{p}}$ ($\gamma_{\mathrm{s}}$) is the corresponding Rabi frequency of the laser-exciton interaction, and $|\mathbf{0}\rangle$ is the vacuum state without exciton. The above Hamiltonian can continuously excite EP pairs with an energy $\omega_{\mathrm{p}} - \omega_{\mathrm{s}}$.

We apply this photoexcitation scheme to the triangular MWC in Fig. 2(b) and set the laser Stokes shift to $\omega_{\mathrm{p}} - \omega_{\mathrm{s}} \approx 2\tilde{\omega}_{1,\mathbf{K}_{\mathrm{m}}}$. The steady state EP occupation, which introduces a spin-dependent geometric energy correction $\langle\hat{H}_{\mathrm{BG}}\rangle$ to MWC, is assumed to be proportional to the photoexcitation rate, whereas the spin configuration of MWC is assumed to be in thermal equilibrium. For a fixed spin temperature and total EP excitation number, we calculate the spin distribution and thermally-averaged magnetic moment self-consistently [41]. The photoexcitation rate of EP is shown in Fig. 4(a), where the generated EP are found to be localized near $\pm\mathbf{K}_{\mathrm{m}}$ and exhibit chirality $\approx 1$ or $-1$. Fig. 4(b) indicates that the thermally-averaged magnetic moment increases when lowering the temperature, and saturates when below a critical value $T_{\mathrm{c}}^{(\mathrm{opt})}$. For a qualitative understanding to this behavior, we approximate the geometric energy correction by $\langle\hat{H}_{\mathrm{BG}}\rangle \approx -\frac{m\Omega}{2}\tau\omega_{1,\mathbf{K}_{\mathrm{m}}}^2 N_{\mathrm{ph}}$ with $N_{\mathrm{ph}}$ the total EP number. The spin distribution $P(\tau) \propto C_{N}^{N(\tau+1)/2}\exp\left(\frac{m\Omega\omega_{1,\mathbf{K}_{\mathrm{m}}}^2 N_{\mathrm{ph}}}{2k_B T}\tau\right)$ then corresponds to a narrow gaussian form centered at $\min\left(\frac{m\Omega\omega_{1,\mathbf{K}_\mathrm{m}}^2}{2k_BT}\frac{N_\mathrm{ph}}{N},1\right)$, see Fig. 4(b) inset. The critical temperature for a full magnetization is $T_\mathrm{c}^{(\mathrm{opt})}\approx\frac{m\Omega\omega_{1,\mathbf{K}_\mathrm{m}}^2}{2k_B}\frac{N_\mathrm{ph}}{N}$, which increases with both the Berry curvature and the EP excitation number.

Applying the same scheme to the honeycomb MWC in Fig. 1(c) can also result in a ferromagnetic phase [41]. Since both MWC and the applied lasers are $\mathcal{P}$-symmetric, it cannot lead to an antiferromagnetic phase which violates $\mathcal{P}$ symmetry. Alternatively, below we consider a $\mathcal{P}$-asymmetric moiré pattern with two inequivalent local potential minima. It can realize a charge-transfer insulator (CTI) in the $\mathcal{P}$-asymmetric honeycomb form under the filling $v = 2$ [52,53]. The EP dispersion for a typical CTI is shown in Fig. 4(c), where all four EP branches near $\pm\mathbf{K}_\mathrm{m}$ are separated by large energy differences. We set the laser Stokes shift to $\omega_\mathrm{p}-\omega_\mathrm{s}\approx\tilde{\omega}_{3,\mathbf{K}_\mathrm{m}}+\tilde{\omega}_{4,-\mathbf{K}_\mathrm{m}}$. The calculated photoexcitation rate in Fig. 4(c) indicates that the generated EP near $-\mathbf{K}_\mathrm{m}$ corresponds to $l=4$ with right-handed (left-handed) circular motions for A (B) sublattice sites, whereas EP near $\mathbf{K}_\mathrm{m}$ corresponds to $l=3$ with left-handed circular motion for A and no motion for B. These chiral EP modes lower the energies of spin configurations with $\tau_\mathrm{A}<0$ and $\tau_\mathrm{B}>0$, resulting in a positive (negative) thermally-averaged magnetic moment for B (A) sublattice, see Fig. 4(d). This then gives rise to a ferrimagnetic phase ($|\tau_\mathrm{A}|\neq|\tau_\mathrm{B}|$), which crosses to antiferromagnetic ($\tau_\mathrm{A}=-\tau_\mathrm{B}$) under a low-enough temperature.

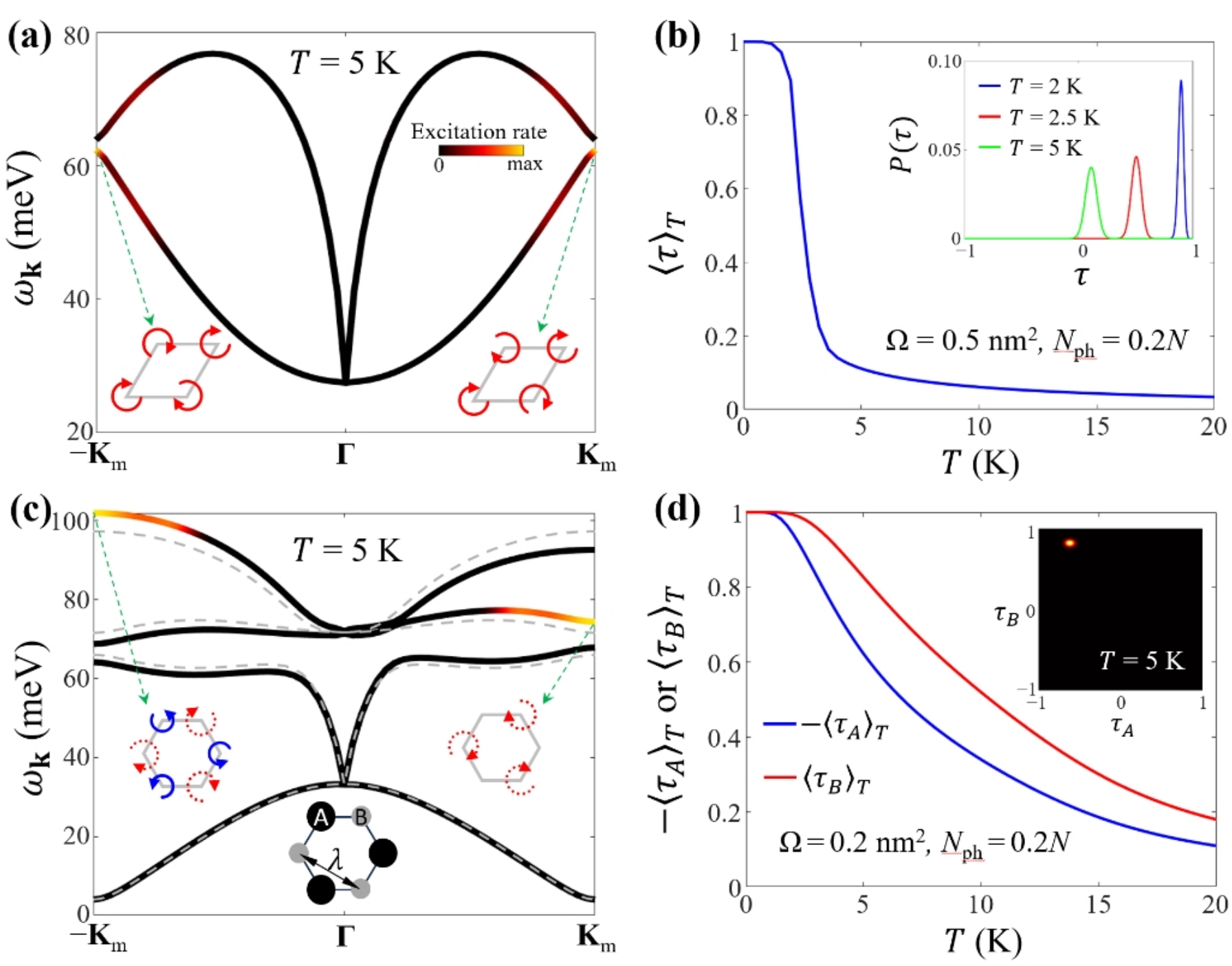

**Figure 4.** (a) The photoexcitation rate of EP in a triangular MWC by the stimulated Raman process with a Stokes shift $2\tilde{\omega}_{1,\mathbf{K}_\mathrm{m}}$. Other parameters are: temperature $T = 5$ K, Berry curvature $\Omega = 0.5\ \mathrm{nm}^2$, total EP number $N_\mathrm{ph} = 0.2N = 80$. A 2 meV linewidth for EP has been assumed. The insets show vibrational

motions of EP modes with peak excitation rates. (b) The thermally-averaged magnetic moment $\langle\tau\rangle_T$ for the triangular MWC. The inset shows the spin distribution function. (c) The EP dispersion of a $\mathcal{P}$-asymmetric charge-transfer insulator (see the lower inset as an illustration), under $\Omega$ = 0.2 nm$^2$, $\lambda$ = 8 nm, $T$ = 5 K and $N_{\mathrm{ph}} = 0.2N = 80$. The moiré confinement strengths are $\gamma_{\mathrm{A}}$ = 5 meV/nm$^2$ for A sublattice and $\gamma_{\mathrm{B}}$ = 10 meV/nm$^2$ for B sublattice. The solid (dashed) lines correspond to the case when the equilibrium magnetism is included (not considered). The color of the solid line shows the photoexcitation rate under a Stokes shift $\tilde{\omega}_{3,\mathbf{K}_{\mathrm{m}}} + \tilde{\omega}_{4,-\mathbf{K}_{\mathrm{m}}}$ and 2 meV EP linewidth. (d) The thermally-averaged magnetic moments $\langle\tau_A\rangle_T$ and $\langle\tau_B\rangle_T$ for the charge-transfer insulator. The inset is the spin distribution $P(\tau_{\mathrm{A}}, \tau_{\mathrm{B}})$ under 5 K.

*Summary and discussion.* We have shown that the electron spin/valley in MWC is coupled to EP through a geometric valley-orbit coupling effect. As a result, magnetism can emerge spontaneously when the temperature is below a critical value, or by a nonequilibrium EP distribution generated through the stimulated Raman process. Depending on the considered systems, out-of-plane ferro-, ferri- and antiferro-magnetic phases can be realized. These results point to an exciting scheme for engineering magnetism through manipulating intrinsic quasiparticles of electron crystals. We expect such a mechanism to be applicable in a series of real materials including the monolayer TMDs, Janus TMDs, multilayer TMDs van der Waals structures, gapped multilayer graphene. For systems without a moiré pattern, an electrostatic potential can be remoted imprinted by the periodic charge distribution in the substrate/capping layer, which has been experimentally realized in monolayer TMDs [54-63], monolayer graphene [64], Bernal bilayer graphene [65-67]. Such electrostatic potentials can facilitate the formation of MWC. Another key factor affecting the experimental observation is the magnitude of the Berry curvature, which determines the critical temperature. In monolayer TMDs, intrinsic Berry curvatures are calculated to be ~ 0.1-0.2 nm$^2$ [40,46], resulting in an extremely small critical temperature ~ $10^{-3}$ meV for the spontaneous magnetization. It increases to an experimentally accessible value ~ 0.1 meV if the Berry curvature can be enhanced to ~ 1 nm$^2$. Since the Berry curvature is inversely proportional to the square of the gap, systems with small gaps can facilitate the realization of EP-induced magnetism, e.g., in gapped multilayer graphene. Meanwhile, the two conduction sub-bands with opposite spins in monolayer TMDs are separated by a small splitting ~ 30 meV [46], where a large Berry curvature can emerge under a Rashba SOC [41]. For example, the internal out-of-plane electric field in Janus TMDs introduces a Rashba parameter as large as ~ 0.2 eV·Å [68], which can give rise to a Berry curvature ~ 1 nm$^2$. In bilayer TMDs systems, the two conduction/valence bands in opposite layers can exhibit a momentum-dependent interlayer hopping [69]. By tuning the interlayer band-offset through an out-of-plane electric field, a large Berry curvature can also be realized [41]. In certain moiré patterns, the spatially-periodic interlayer coupling can result in giant Berry curvatures due to the small energy separations (several meV) between moiré mini-bands [70,71], where our perturbative treatment for the Berry phase effect does not apply.

Furthermore, the strong electron-electron interaction in MWC can efficiently mix different mini-band. In this case, the Abelian Berry curvature of a single band becomes inappropriate, and the quantitative result needs a non-Abelian model which takes into account the band-mixing effect.

*Acknowledgments.* H.Y. acknowledges support by NSFC under Grant No. 12274477 and Guangdong Provincial Quantum Science Strategic Initiative (GDZX2501003).

[1] Kin Fai Mak and Jie Shan, *Semiconductor moiré materials*, Nat. Nanotech. **17**, 686 (2022).
[2] Nathan P. Wilson, Wang Yao, Jie Shan, and Xiaodong Xu, *Excitons and emergent quantum phenomena in stacked 2D semiconductors*, Nature **599**, 383 (2021).
[3] Yanhao Tang, Lizhong Li, Tingxin Li, Yang Xu, Song Liu, Katayun Barmak, Kenji Watanabe, Takashi Taniguchi, Allan H. MacDonald, Jie Shan *et al.*, *Simulation of Hubbard model physics in WSe2/WS2 moiré superlattices*, Nature **579**, 353 (2020).
[4] Emma C. Regan, Danqing Wang, Chenhao Jin, M. Iqbal Bakti Utama, Beini Gao, Xin Wei, Sihan Zhao, Wenyu Zhao, Zuocheng Zhang, Kentaro Yumigeta *et al.*, *Mott and generalized Wigner crystal states in WSe2/WS2 moiré superlattices*, Nature **579**, 359 (2020).
[5] Yuya Shimazaki, Ido Schwartz, Kenji Watanabe, Takashi Taniguchi, Martin Kroner, and Ataç Imamoğlu, *Strongly correlated electrons and hybrid excitons in a moiré heterostructure*, Nature **580**, 472 (2020).
[6] Lei Wang, En-Min Shih, Augusto Ghiotto, Lede Xian, Daniel A. Rhodes, Cheng Tan, Martin Claassen, Dante M. Kennes, Yusong Bai, Bumho Kim *et al.*, *Correlated electronic phases in twisted bilayer transition metal dichalcogenides*, Nat. Mater. **19**, 861 (2020).
[7] Zhaodong Chu, Emma C Regan, Xuejian Ma, Danqing Wang, Zifan Xu, M. Iqbal Bakti Utama, Kentaro Yumigeta, Mark Blei, Kenji Watanabe, Takashi Taniguchi *et al.*, *Nanoscale Conductivity Imaging of Correlated Electronic States in WSe2/WS2 Moiré Superlattices*, Phys. Rev. Lett. **125**, 186803 (2020).
[8] Yang Xu, Song Liu, Daniel A Rhodes, Kenji Watanabe, Takashi Taniguchi, James Hone, Veit Elser, Kin Fai Mak, and Jie Shan, *Correlated insulating states at fractional fillings of moiré superlattices*, Nature **587**, 214 (2020).
[9] Xiong Huang, Tianmeng Wang, Shengnan Miao, Chong Wang, Zhipeng Li, Zhen Lian, Takashi Taniguchi, Kenji Watanabe, Satoshi Okamoto, Di Xiao *et al.*, *Correlated insulating states at fractional fillings of the WS2/WSe2 moiré lattice*, Nat. Phys. **17**, 715 (2021).
[10] Erfu Liu, Takashi Taniguchi, Kenji Watanabe, Nathaniel M. Gabor, Yong-Tao Cui, and Chun Hung Lui, *Excitonic and Valley-Polarization Signatures of Fractional Correlated Electronic Phases in a WSe2/WS2 Moiré Superlattice*, Phys. Rev. Lett. **127**, 037402 (2021).
[11] Hongyuan Li, Shaowei Li, Emma C. Regan, Danqing Wang, Wenyu Zhao, Salman Kahn, Kentaro Yumigeta, Mark Blei, Takashi Taniguchi, Kenji Watanabe *et al.*, *Imaging two-dimensional generalized Wigner crystals*, Nature **597**, 650 (2021).
[12] Hongyuan Li, Shaowei Li, Mit H. Naik, Jingxu Xie, Xinyu Li, Emma Regan, Danqing Wang, Wenyu Zhao, Kentaro Yumigeta, Mark Blei *et al.*, *Imaging local discharge cascades for correlated electrons in WS2/WSe2 moiré superlattices*, Nat. Phys. **17**, 1114 (2021).
[13] Yuya Shimazaki, Clemens Kuhlenkamp, Ido Schwartz, Tomasz Smoleński, Kenji Watanabe, Takashi Taniguchi, Martin Kroner, Richard Schmidt, Michael Knap, and Ataç Imamoğlu, *Optical Signatures of Periodic Charge Distribution in a Mott-like Correlated Insulator State*, Phys. Rev. X **11**, 021027 (2021).
[14] Chenhao Jin, Zui Tao, Tingxin Li, Yang Xu, Yanhao Tang, Jiacheng Zhu, Song Liu, Kenji Watanabe, Takashi Taniguchi, James C. Hone *et al.*, *Stripe phases in WSe2/WS2 moiré superlattices*, Nat. Mater. **20**, 940 (2021).
[15] Tomasz Smoleński, Pavel E. Dolgirev, Clemens Kuhlenkamp, Alexander Popert, Yuya Shimazaki, Patrick Back, Xiaobo Lu, Martin Kroner, Kenji Watanabe, Takashi Taniguchi *et al.*, *Signatures of Wigner crystal of electrons in a monolayer semiconductor*, Nature **595**, 53 (2021).
[16] You Zhou, Jiho Sung, Elise Brutschea, Ilya Esterlis, Yao Wang, Giovanni Scuri, Ryan J. Gelly, Hoseok Heo, Takashi Taniguchi, Kenji Watanabe *et al.*, *Bilayer Wigner crystals in a transition metal dichalcogenide heterostructure*, Nature **595**, 48 (2021).

[17] Fengcheng Wu, Timothy Lovorn, Emanuel Tutuc, and A. H. MacDonald, *Hubbard Model Physics in Transition Metal Dichalcogenide Moiré Bands*, Phys. Rev. Lett. **121**, 026402 (2018).

[18] Nitin Kaushal, Nicolás Morales-Durán, Allan H. MacDonald, and Elbio Dagotto, *Magnetic ground states of honeycomb lattice Wigner crystals*, Commun. Phys. **5**, 289 (2022).

[19] Nicolás Morales-Durán, Pawel Potasz, and Allan H. MacDonald, *Magnetism and quantum melting in moiré-material Wigner crystals*, Phys. Rev. B **107**, 235131 (2023).

[20] Nai Chao Hu and Allan H. MacDonald, *Competing magnetic states in transition metal dichalcogenide moiré materials*, Phys. Rev. B **104**, 214403 (2021).

[21] Jiawei Zang, Jie Wang, Jennifer Cano, and Andrew J. Millis, *Hartree-Fock study of the moiré Hubbard model for twisted bilayer transition metal dichalcogenides*, Phys. Rev. B **104**, 075150 (2021).

[22] Haining Pan, Fengcheng Wu, and Sankar Das Sarma, *Quantum phase diagram of a Moiré-Hubbard model*, Phys. Rev. B **102**, 201104(R) (2020).

[23] Xi Wang, Chengxin Xiao, Heonjoon Park, Jiayi Zhu, Chong Wang, Takashi Taniguchi, Kenji Watanabe, Jiaqiang Yan, Di Xiao, Daniel R. Gamelin *et al.*, *Light-induced ferromagnetism in moiré superlattices*, Nature **604**, 468 (2022).

[24] Chengxin Xiao, Yong Wang, and Wang Yao, *Dynamic Generation of Spin Spirals of Moiré Trapped Carriers via Exciton Mediated Spin Interactions*, Nano Lett. **23**, 1872 (2023).

[25] Hui Yang and Ya-Hui Zhang, *Exciton- and light-induced ferromagnetism from doping a moiré Mott insulator*, Phys. Rev. B **110**, L041115 (2024).

[26] Urban F. P. Seifert and Leon Balents, *Spin Polarons and Ferromagnetism in Doped Dilute Moiré-Mott Insulators*, Phys. Rev. Lett. **132**, 046501 (2024).

[27] L. Ciorciaro, T. Smoleński, I. Morera, N. Kiper, S. Hiestand, M. Kroner, Y. Zhang, K. Watanabe, T. Taniguchi, E. Demler *et al.*, *Kinetic magnetism in triangular moiré materials*, Nature **623**, 509 (2023).

[28] Yanhao Tang, Kaixiang Su, Lizhong Li, Yang Xu, Song Liu, Kenji Watanabe, Takashi Taniguchi, James Hone, Chao-Ming Jian, Cenke Xu *et al.*, *Evidence of frustrated magnetic interactions in a Wigner-Mott insulator*, Nat. Nanotech. **18**, 233 (2023).

[29] Wenjin Zhao, Bowen Shen, Zui Tao, Sunghoon Kim, Patrick Knüppel, Zhongdong Han, Yichi Zhang, Kenji Watanabe, Takashi Taniguchi, Debanjan Chowdhury *et al.*, *Emergence of ferromagnetism at the onset of moiré Kondo breakdown*, Nat. Phys. **20**, 1772 (2024).

[30] Lynn Bonsall and A. A. Maradudin, *Some static and dynamical properties of a two-dimensional Wigner crystal*, Phys. Rev. B **15**, 1959 (1977).

[31] R. Côté and A. H. MacDonald, *Collective modes of the two-dimensional Wigner crystal in a strong magnetic field*, Phys. Rev. B **44**, 8759 (1991).

[32] Hongyi Yu and Jiyong Zhou, *Phonons of electronic crystals in two-dimensional semiconductor moiré patterns*, Nat. Sci. **3**, e20220065 (2023).

[33] Samuel Brem and Ermin Malic, *Terahertz Fingerprint of Monolayer Wigner Crystals*, Nano Lett. **22**, 1311 (2022).

[34] Junkai Dong, Ophelia Evelyn Sommer, Tomohiro Soejima, Daniel E. Parker, and Ashvin Vishwanath, *Phonons in electron crystals with Berry curvature*, PNAS **122**, e2515532122 (2025).

[35] Joonho Jang, Benjamin M. Hunt, Loren N. Pfeiffer, Kenneth W. West, and Raymond C. Ashoori, *Sharp tunnelling resonance from the vibrations of an electronic Wigner crystal*, Nat. Phys. **13**, 340 (2017).

[36] Lili Zhao, Wenlu Lin, Yoon Jang Chung, Adbhut Gupta, Kirk W. Baldwin, Loren N. Pfeiffer, and Yang Liu, *Dynamic Response of Wigner Crystals*, Phys. Rev. Lett. **130**, 246401 (2023).

[37] L. Wang, F. Menzel, F. Pichler, P. Knüppel, K. Watanabe, T. Taniguchi, M. Knap, and T. Smoleński, *Spectroscopy of Wigner crystal polarons in an atomically thin semiconductor*, arXiv:2512.16552.

[38] Lifu Zhang, Liuxin Gu, Haydn S. Adlong, Arthur Christianen, Eugen Dizer, Ruihao Ni, Rundong Ma, Suji Park, Houk Jang, Takashi Taniguchi *et al.*, *Wigner polarons reveal Wigner crystal dynamics in a monolayer semiconductor*, arXiv:2512.16631.

[39] Yan Zhao, Yuhang Hou, Xiangbin Cai, Shihao Ru, Shunshun Yang, Yan Zhang, Xuran Dai, Qiuyu Shang, Abdullah Rasmita, Haiyang Pan *et al.*, *Electronic Phonons in a Moiré Electron Crystal*, arXiv:2512.18217.

[40] Di Xiao, Gui-Bin Liu, Wanxiang Feng, Xiaodong Xu, and Wang Yao, *Coupled Spin and Valley Physics in Monolayers of MoS2 and Other Group-VI Dichalcogenides*, Phys. Rev. Lett. **108**, 196802 (2012).

[41] *See the Supplemental Material for details.*

[42] Ajit Srivastava and Ataç Imamoğlu, *Signatures of Bloch-Band Geometry on Excitons: Nonhydrogenic Spectra in Transition-Metal Dichalcogenides*, Phys. Rev. Lett. **115**, 166802 (2015).

[43] Jianhui Zhou, Wen-Yu Shan, Wang Yao, and Di Xiao, *Berry Phase Modification to the Energy Spectrum of Excitons*, Phys. Rev. Lett. **115**, 166803 (2015).
[44] Jianju Tang, Songlei Wang, and Hongyi Yu, *Inheritance of the exciton geometric structure from bloch electrons in two-dimensional layered semiconductors*, Front. Phys. **19**, 43210 (2024).
[45] Chaw-Keong Yong, M. Iqbal Bakti Utama, Chin Shen Ong, Ting Cao, Emma C. Regan, Jason Horng, Yuxia Shen, Hui Cai, Kenji Watanabe, Takashi Taniguchi *et al.*, *Valley-dependent exciton fine structure and Autler-Townes doublets from Berry phases in monolayer MoSe2*, Nat. Mater. **18**, 1065 (2019).
[46] Gui-Bin Liu, Di Xiao, Yugui Yao, Xiaodong Xu, and Wang Yao, *Electronic structures and theoretical modelling of two-dimensional group-VIB transition metal dichalcogenides*, Chem. Soc. Rev. **44**, 2643 (2015).
[47] Di Xiao, Wang Yao, and Qian Niu, *Valley-Contrasting Physics in Graphene: Magnetic Moment and Topological Transport*, Phys. Rev. Lett. **99**, 236809 (2007).
[48] William Holtzmann, Weijie Li, Eric Anderson, Jiaqi Cai, Heonjoon Park, Chaowei Hu, Takashi Taniguchi, Kenji Watanabe, Jiun-Haw Chu, Di Xiao *et al.*, *Optical control of integer and fractional Chern insulators*, Nature **649**, 1147 (2026).
[49] O. Huber, K. Kuhlbrodt, E. Anderson, W. Li, K. Watanabe, T. Taniguchi, M. Kroner, X. Xu, A. Imamoğlu, and T. Smoleński, *Optical control over topological Chern number in moiré materials*, Nature **649**, 1153 (2026).
[50] Xiangbin Cai, Haiyang Pan, Yuzhu Wang, Abdullah Rasmita, Shunshun Yang, Yan Zhao, Wei Wang, Ruihuan Duan, Ruihua He, Kenji Watanabe *et al.*, *Optical switching of a moiré Chern ferromagnet*, Nature **650**, 580 (2026).
[51] Fabian Pichler, Clemens Kuhlenkamp, and Michael Knap, *False Vacuum Decay in Flat-Band Ferromagnets: Role of Quantum Geometry and Chiral Edge States*, Phys. Rev. Lett. **136**, 156502 (2026).
[52] Yang Zhang, Noah F. Q. Yuan, and Liang Fu, *Moiré quantum chemistry: Charge transfer in transition metal dichalcogenide superlattices*, Phys. Rev. B **102**, 201115(R) (2020).
[53] Zhexu Shan, Wenjian Su, Kenji Watanabe, Takashi Taniguchi, Shiyao Zhu, Yuanfeng Xu, and Yanhao Tang, *Gate-tunable chiral spin mode in WSe2/WS2 moiré superlattices*, arXiv:2509.16511.
[54] Yang Xu, Connor Horn, Jiacheng Zhu, Yanhao Tang, Liguo Ma, Lizhong Li, Song Liu, Kenji Watanabe, Takashi Taniguchi, James C. Hone *et al.*, *Creation of moiré bands in a monolayer semiconductor by spatially periodic dielectric screening*, Nat. Mater. **20**, 645 (2021).
[55] Qianying Hu, Zhen Zhan, Huiying Cui, Yalei Zhang, Feng Jin, Xuan Zhao, Mingjie Zhang, Zhichuan Wang, Qingming Zhang, Kenji Watanabe *et al.*, *Observation of Rydberg moiré excitons*, Science **380**, 1367 (2023).
[56] Dong Seob Kim, Roy C. Dominguez, Rigo Mayorga-Luna, Dingyi Ye, Jacob Embley, Tixuan Tan, Yue Ni, Zhida Liu, Mitchell Ford, Frank Y. Gao *et al.*, *Electrostatic moiré potential from twisted hexagonal boron nitride layers*, Nat. Mater. **23**, 65 (2024).
[57] Jie Gu, Jiacheng Zhu, Patrick Knuppel, Kenji Watanabe, Takashi Taniguchi, Jie Shan, and Kin Fai Mak, *Remote imprinting of moiré lattices*, Nat. Mater. **23**, 219 (2024).
[58] Minhao He, Jiaqi Cai, Huiyuan Zheng, Eric Seewald, Takashi Taniguchi, Kenji Watanabe, Jiaqiang Yan, Matthew Yankowitz, Abhay Pasupathy, Wang Yao *et al.*, *Dynamically tunable moiré exciton Rydberg states in a monolayer semiconductor on twisted bilayer graphene*, Nat. Mater. **23**, 224 (2024).
[59] Liuxin Gu, Lifu Zhang, Sam Felsenfeld, Beini Gao, Rundong Ma, Suji Park, Houk Jang, Takashi Taniguchi, Kenji Watanabe, and You Zhou, *Quantum Confining Excitons with an Electrostatic Moire´ Superlattice*, Phys. Rev. Lett. **135**, 026901 (2025).
[60] Natasha Kiper, Haydn S. Adlong, Arthur Christianen, Martin Kroner, Kenji Watanabe, Takashi Taniguchi, and Atac İmamoğlu, *Confined Trions and Mott-Wigner States in a Purely Electrostatic Moiré Potential*, Phys. Rev. X **15**, 011049 (2025).
[61] Minhyun Cho, Biswajit Datta, Kwanghee Han, Saroj B. Chand, Pratap Chandra Adak, Sichao Yu, Fengping Li, Kenji Watanabe, Takashi Taniguchi, James Hone *et al.*, *Moiré Exciton Polaron Engineering via twisted hBN*, Nano Lett. **25**, 1381 (2025).
[62] Dong Seob Kim, Chengxin Xiao, Roy C. Dominguez, Zhida Liu, Hamza Abudayyeh, Kyoungpyo Lee, Rigo Mayorga- Luna, Hyunsue Kim, Kenji Watanabe, Takashi Taniguchi *et al.*, *Moiré ferroelectricity modulates light emission from a semiconductor monolayer*, Sci. Adv. **11**, eadt7789 (2025).
[63] Amine Ben Mhenni, Elif Çetiner, Kenji Watanabe, Takashi Taniguchi, Jonathan J. Finley, and Nathan P. Wilson, *Engineering strong correlations in a perfectly aligned dual moiré system*, arXiv:2509.13159.
[64] Zi-Han Guo, Chao Yan, Jia-Qi He, Ke Lv, Kenji Watanabe, Takashi Taniguchi, Ya-Ning Ren,

and Lin He, *Manipulating Superlattice Potentials and Quantum Confinement in Graphene via Moiré Ferroelectricity*, Nano Lett. **25**, 11543 (2025).
[65] Jing Ding, Hanxiao Xiang, Wenqiang Zhou, Naitian Liu, Qianmei Chen, Xinjie Fang, Kangyu Wang, Linfeng Wu, Kenji Watanabe, Takashi Taniguchi *et al.*, *Engineering band structures of two-dimensional materials with remote moiré ferroelectricity*, Nat. Commun. **15**, 9087 (2024).
[66] Zuocheng Zhang, Jingxu Xie, Wenyu Zhao, Ruishi Qi, Collin Sanborn, Shaoxin Wang, Salman Kahn, Kenji Watanabe, Takashi Taniguchi, Alex Zettl *et al.*, *Engineering correlated insulators in bilayer graphene with a remote Coulomb superlattice*, Nat. Mater. **23**, 189 (2024).
[67] Xirui Wang, Cheng Xu, Samuel Aronson, Daniel Bennett, Nisarga Paul, Philip J. D. Crowley, Clément Collignon, Kenji Watanabe, Takashi Taniguchi, Raymond Ashoori *et al.*, *Moiré band structure engineering using a twisted boron nitride substrate*, Nat. Commun. **16**, 178 (2025).
[68] Tao Hu, Fanhao Jia, Guodong Zhao, Jiongyao Wu, Alessandro Stroppa, and Wei Ren, *Intrinsic and anisotropic Rashba spin splitting in Janus transition-metal dichalcogenide monolayers*, Phys. Rev. B **97**, 235404 (2018).
[69] Qingjun Tong, Hongyi Yu, Qizhong Zhu, Yong Wang, Xiaodong Xu, and Wang Yao, *Topological mosaics in moiré superlattices of van der Waals heterobilayers*, Nat. Phys. **13**, 356 (2017).
[70] Fengcheng Wu, Timothy Lovorn, Emanuel Tutuc, Ivar Martin, and A. H. MacDonald, *Topological Insulators in Twisted Transition Metal Dichalcogenide Homobilayers*, Phys. Rev. Lett. **122**, 086402 (2019).
[71] Hongyi Yu, Mingxing Chen, and Wang Yao, *Giant magnetic field from moiré induced Berry phase in homobilayer semiconductors*, Natl. Sci. Rev. **7**, 12 (2020).